\documentclass[titlepage,twocolumn,amsmath,amssymb,floatfix,superscriptaddress]{revtex4}

\usepackage{graphicx}
\usepackage{amsmath,amssymb}
\usepackage[T1]{fontenc}
\usepackage[latin1]{inputenc}
\usepackage{times}
\usepackage{bm}
\usepackage[usenames,dvips]{color}
\usepackage[normalem]{ulem}
\usepackage{setspace}
  \makeatletter
  \def\@dotsep{4.5}
  \makeatother
\newlength{\myVSpace}
\newcommand{\Refon}[1]{Ref.~\onlinecite{#1}}

\newcommand{\T}{\langle T\rangle}
\newcommand{\U}{\langle U\rangle}

\begin{document}
\title{Pinning effect and QPT-like behavior for two particles confined by a
core-shell potential}
\author{P.P. Marchisio
}
\email{ppm501@york.ac.uk}
\affiliation{
Department of Physics, University of York, York YO10 5DD, United
Kingdom.
}
\author{J.P. Coe
}
\email{jpc503@york.ac.uk}
\affiliation{
Department of Physics, University of York, York YO10 5DD, United
Kingdom.
}
\affiliation{Department of Chemistry, School of Engineering and Physical Sciences,
Heriot-Watt University, Edinburgh, EH14 4AS, UK}
\author{I. D'Amico
}
\email{irene.damico@york.ac.uk}
\affiliation{
Department of Physics, University of York, York YO10 5DD, United
Kingdom.
}

\date{today}

\begin{abstract}
We study the ground state entanglement, energy and fidelities of a two-electron
system  bounded by  a core-shell potential, where the core width is varied  continuously
 until it eventually vanishes. This simple system displays a rich and
complex behavior: as the core width is varied, this system is characterized by
two peculiar transitions where, for different reasons, it displays
characteristics similar to a few-particle quantum phase transition. The first
occurrence corresponds to something akin to a second order quantum phase
transition, while the second transition is marked by  a discontinuity,
with respect to the driving parameter, in the first derivatives of quantities
like energy and entanglement. The study of this system allows to shed light on
the sudden variation of entanglement and energy  observed in
\Refon{Abdullah:2009:PRB}.
We also compare the core-shell system with a system where a core well is
absent: this shows that, even when extremely narrow, the core well has  a
relevant `pinning' effect. Interestingly, depending on the potential symmetry, the pinning of the wavefunction may either   halve or double the system  entanglement (with respect to the
no-core-well system) when
the ground state is already bounded to the outer (shell) well.
In the process we discuss the system fidelity and show the usefulness of
considering the particle density fidelity as opposed to the more commonly
used -- but much more difficult to access -- wavefunction fidelity. In
particular we demonstrate that -- for ground-states with nodeless spatial wavefunctions -- the  particle density
fidelity is zero if and only if the wavefunction fidelity is zero.
\end{abstract}

\maketitle

\section{Introduction}

The realization of the importance of entanglement  triggered a rethink in the
way one can understand and quantify some quantum processes. Indeed, quantum
information theory (QIT)  has stemmed from the application of entanglement and
the superposition principle to the processing and transmission of
data,\cite{nielsen} and   it is now acknowledged that entanglement can play a
central role in the description and understanding  of   quantum phase
transitions (QPTs).\cite{Larsson:2006:PRA,Larsson:2005:PRL,Chen:2006:NJP} In
QIT and QPTs it is important to determine how a quantum state changes under
quantum operations or by  varying external parameters.    The fidelity
\cite{anandan:1991:JFP,nielsen} -- extensively used in QIT to assess the
`closeness' of different quantum states -- may naturally encompass the effect
of a driving parameter on a system, and, as such,  it has been proposed as a
key tool in understanding
QPTs.\cite{zanardi:2006:PRE,zanardi:2007:PRL,You:2007:PRE} Entanglement and
fidelity can then provide  a common language for QIT and
QPTs.\cite{amin:2009:PRA,rezakhani:2010:arxiv,quan:2006:PRL} The definition of a QPT has been widened by some authors to include changes in
the quantum state of few-particle systems such as singlet-triplet transitions
in a single quantum dot.\cite{Roch:2009:nature} Few-particle systems have also been used to characterize the predictive power of QPT indicators  for a system undergoing a QPT in the thermodynamical limit.\cite{Oh:2009:PLA}

In previous work, \cite{Abdullah:2009:PRB} it was shown that the  transition
from a core-shell to a double well potential induces a sudden variation of both
the entanglement and the energy of two  electrons initially confined  within
the core well. This variation becomes sharper as the confining potential
becomes harder, i.e. more similar to a rectangular-like potentials. This steep
variation was regarded as something potentially akin to a QPT but in 
the few-particle case.

In order to  understand this phenomenon,   in this paper we will study  systems
related to \Refon{Abdullah:2009:PRB} and characterized by  {\it
rectangular-like} confining potential. We will focus on  how ground-state
entanglement, energy, and  fidelities are affected by varying the potential
core width and show that these simple  systems encompass indeed a rich and
complex behavior. The system we will mainly concentrate on is given by two electrons trapped
within a core-shell  potential, whose core reduces in width until it eventually
disappears  (see Fig.~\ref{potentialfig}). This may represent a (core-shell)
quantum dot  with an externally-driven confining  potential: quantum dots are
one of the most promising hardware for the physical realization of QIT
devices,\cite{loss:1998:PRA,burkard:1999:PRB,Reina:2000:PRA,Biolatti:2002:PRB,Li:2003:science,Pazy:2003:EL,
Feng:2004:EL,Hodgson:2007:JAP,Spiller:2009:NJP} hence, our findings may be of
interest for QIT applications. The system ground state is initially bound to  the core well, but will become
bound to the outer well (or shell) as the core width is reduced to zero and the
outer well width increases. We will show that the corresponding sharp
entanglement variation is  characterized by {\it two very different
transitions}. The first presents elements  akin to  a second-order QPT and is associated with the
transition of the ground state from the core  to the outer well; the second is
marked by a {\it discontinuity} in the energy and entanglement {\it
derivatives} with respect to the driving parameter,  and we demonstrate that it is
due to the peculiarities of the confining potential. We will also explicitly
discuss the implications of these findings for the system described in
\Refon{Abdullah:2009:PRB}.

Our analysis is important in the context of local sensitivity analysis. In particular, 
due to the pivotal role that the entanglement plays in several\cite{note:Sen:Anal} 
quantum protocols (such as quantum algorithms\cite{Jozsa:2003:PRSL},  quantum teleportation\cite{Bennett:1993:PRL} and some quantum cryptography protocols\cite{Heiss:2002:book}) here we report on the sensitivity of the 
entanglement with respect to  small variations of the
external parameter,\cite{Rabitz:1983:ARPC,Swillam:2008:OpComm}   
characterize the region of the parameter space over which the entanglement 
shows the  steepest variation and, consequently, ascertain the possibility of employing the potential variations as entanglement `switch'. 
In fact our calculations show that the presence of an inner core has a {\it
strong pinning effect} on the entanglement  even when the ground state is
already bound to the outer well. Depending on the system geometry, it may in fact either  halve  the entanglement (symmetric system) or double its value (asymmetric systems) when  compared to the corresponding system without  core well.
This might potentially be exploited  to induce sharp variations (switch)
of the entanglement by modifying small regions of the confining potential.

Finally, in the spirit of density-functional theory,\cite{hohenberg:1964:PR} we will
study whether the particle density can be used to track the system's ground state
behavior via a {\it particle-density fidelity}.  We note that, from an
experimental point of view, the density is  a more accessible quantity than the
full system wavefunction; our results show that, at least for the system at
hand, the particle-density fidelity delivers  information  similar to the
wavefunction fidelity. Importantly we will demonstrate that, for ground-states with nodeless spatial wavefunctions, {\it the
particle-density fidelity is zero if and only if the wavefunction fidelity is
zero}.

\section{Symmetric potential, model systems\label{sec_models}}
We will first concentrate on systems with a symmetric confining potential (see Fig.~\ref{potentialfig}).

We consider three one-dimensional systems, each consisting of two interacting
electrons bound by a confining potential and whose Hamiltonian in effective
atomic units is
\begin{equation}
H=\sum_{j=1}^{2}\left[ -\frac{1}{2}\frac{\partial^{2}}{\partial
x^{2}_{j}}+V_{i}(x_{j},R) \right] + U\left(x_1,x_2\right).
\label{hamiltonian}
\end{equation}
Here  we  set $U\left(x_1,x_2\right)=\delta(x_1-x_2)$ to represent a contact
Coulomb repulsion between the electrons. $V_{i}(x_{j},R)$ are the confining
potentials characterizing the three systems, $i=DIW,OWO\text{ and }DW$, see
below.

\subsection{System with a `disappearing' inner well} The potential of the
`disappearing' inner well (DIW) system, $V_{DIW}(x;R)$, is characterized by an
inner (core) and an outer shell well, see Fig.~\ref{potentialfig}. As the parameter
$R$ increases, the inner well width, $W^{iw}$, becomes narrower and the outer
well width, $W^{ow}$,  increases as
\begin{align}
& W^{iw}(R)=\left\{\begin{array}{ll}
 w-R &\text{for } R< w \label{Wiw}\\
 0, & \text{for } R \ge w \end{array}\right.\\
 & W^{ow}(R) =w+R.
\end{align}
Taking $V_0$ as the depth of the outer well, we can write

\begin{equation}
V_{DIW}(x;R<w) =\left\{
 \begin{array}{ll}
 2V_0 \!\! & \text{for }\left|\frac{W^{iw}}{2}\right| > |x| \\
  V_0 \!\! & \text{for } \left|\frac{W^{ow}}{2}\right| > |x| \ge
\left|\frac{W^{iw}}{2}\right|\\
 0 \!\! & \text{otherwise}
\end{array}\right.\label{pot_na1}
\end{equation}
and
\begin{equation}
 V_{DIW}(x;R\ge w) =\left\{
 \begin{array}{ll}
  V_0  & \text{for } \left|\frac{W^{ow}}{2}\right| > |x| \\
 0 & \text{otherwise}.
\end{array}\right.\label{pot_na2}
\end{equation}

$V_{DIW}$ has a compact representation through the Heaviside step function,

\begin{equation}
V_{DIW}(x;R)=  V^{iw}(x;R)+V^{ow}(x;R),
\label{analyticalpot}
\end{equation}
where
\begin{equation}
V^{iw}(x;R)\equiv
V_0\left[\theta\left(x+(w-R)/2\right)\theta\left(-x+(w-R)/2\right)\right]
\end{equation} and $V^{ow}(x;R)=V^{iw}(x;-R)$
describe the inner and the outer well, respectively. Eq.~(\ref{analyticalpot})
is equivalent to Eqs.~(\ref{pot_na1}) and (\ref{pot_na2}) if we assign
$\theta(0)=0$.
This is consistent with considering the Heaviside step function $\theta(x)$
as, for example,  the limit (in a distribution
sense~\cite{GeneralisedFunction:book}) for $p\rightarrow \infty$ of
\begin{equation}
\theta_{p}\left(x\right)=\frac{1-e^{-\left(px\right)^2}}{1+e^{-mpx}},
\label{theta}
\end{equation}
where $p$ and $m$ are positive integers. With $p\sim 10$ and $m\sim 20$, we get
a smooth, `softer' version of $V_{DIW}$.    As $p\rightarrow \infty$ arguments
similar to the ones developed in \Refon{Abdullah:2009:PRB} seem to suggest a
discontinuity in the entanglement entropy and energy derivatives (and hence
something reminiscent of a QPT in the few-particle regime). The chosen parametrization
for the potential will help us to better understand this limit.

\subsection{Benchmark system}

\begin{figure}[ht]
\centering
  \includegraphics[width=.4\textwidth]{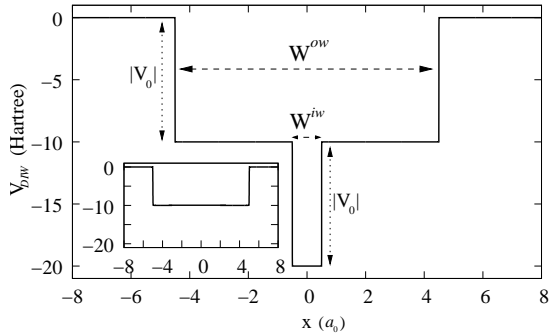}
\caption{Potential $V_{DIW}$ versus $x$ for $R=4\,a_0$. Inset: same as main
panel but for  $R=5\,a_0$, for which DIW and OWO systems coincide.}
\label{potentialfig}
\end{figure}

The confining potential of the `outer well only' (OWO) system  is given by
$V_{OWO}\equiv V^{ow}$, see inset of Fig.~\ref{potentialfig}. We use this
system as a benchmark.

\subsection{Core-shell to double well system}

This is the rectangular-like limit of the system considered in \Refon{Abdullah:2009:PRB}.
As the driving parameter changes, this potential is modified from a core-shell
to a double-well potential. The explicit expression of this potential in the
rectangular-like limit   can be written as
\begin{eqnarray}
\nonumber V_{DW}(x;R)=V_{0}\big[\theta(-x+(w-R)/2)\theta(x+(3w-R)/2)\\
+\theta(-x+(3w-R)/2)\theta(x+(w-R)/2)\big].
\label{eq:doublewell}
\end{eqnarray}
Here the transformations $R_{DW}=2w - R$ and $d=w/2$ give the control parameter
and  the inter-well distance as used  in \Refon{Abdullah:2009:PRB},
respectively.

For the subsequent calculations, unless otherwise stated, we use   $w = 5$
$a_0$, where $a_0$ is the Bohr radius, and $V_0= - 10$ Hartree.

\section{Results for Entanglement and Energy (DIW and OWO systems)}

To calculate the ground-state properties, we directly diagonalize the
Hamiltonian Eq.~(\ref{hamiltonian}), by writing its eigenfunctions $\Psi_{k}$
as a linear combination of single-particle basis functions and truncating the
corresponding expansion as

\begin{equation}
\Psi_{k}(x_1,x_2)=\sum_{j_{1}=1}^{M}\sum_{j_{2}=1}^{M}
a_{j_{1},j_{2};k}\eta_{_{j_{1}}}(x_1;\omega)\eta_{_{j_{2}}}(x_2;\omega),
\label{eigenfunction}
\end{equation}
where $\eta_{_{j}}(x;\omega)$ are the eigenfunctions of the one-dimensional
harmonic oscillator with angular frequency $\omega$. A single-particle basis
size of $M=50$ with $\omega=2$ ensures good convergence of the results at any
$R$.

We calculate the particle-particle spatial entanglement\cite{Coe:2008:PRB}  and
the ground-state energy of the system for $4\,a_0\le R\le8\,a_0$. For the DIW
system $R=4\,a_0$ corresponds to a core-shell structure with the two electrons
confined in the inner well, while for $R\ge 5\,a_0$ we have  $V_{DIW}=V_{OWO}$.

\subsection{Energy}

\begin{figure}[ht!]
\centering
  \includegraphics[width=.4\textwidth]{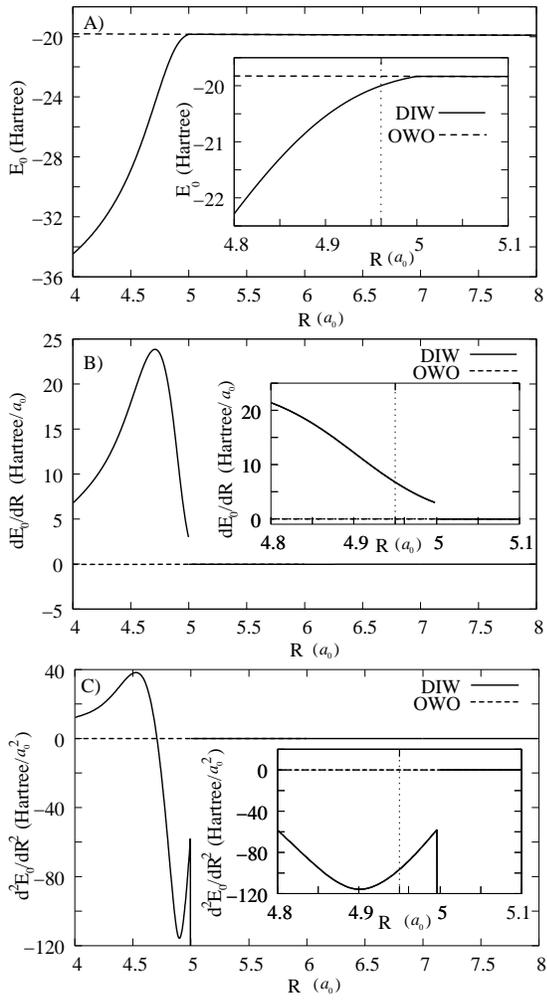}
\caption{ Ground state energy $E_0$ (panel A), first and second derivative of
$E_0$ with respect to $R$ (panel B and C, respectively) for the DIW (solid
line) and OWO (dashed line)  potentials as functions of the driving parameter
$R$. In all the three panels the inset zooms on the `migration region' with
$R_c=4.96\,a_0$ indicated by a vertical dotted line.}
\label{energiesFig}
\end{figure}

First we consider the ground state energy $E_0$ of the DIW system  (solid line
in Fig.~\ref{energiesFig}A) against the benchmark (OWO, dashed line).

As $R$ becomes larger, the inner well narrows  and   the energy   of
the two-electron  state increases, until the electrons are eventually `forced' into the
outer well. The ground state energy leaves the inner well at  $R\equiv R_c
=4.96\,a_0$. This corresponds to an  inner to outer well ratio of  $0.0039$.
Hereafter, we will refer to the parameter region around $R_c$ as  the
`migration region': for these values of the driving parameter the system wavefunction
is the most sensitive to
driving parameter changes. Here 
the electron wavefunction `expands' into the outer well and, as a
consequence of this, the system shows the most interesting behavior.
This region of high sensitivity is relatively narrow and in fact for $R\geqslant 5\,a_0$ 
the  ground state energy becomes a very
slowly-varying, decreasing function of  $R$.

The first derivative of the ground state energy  with respect to the driving
parameter, $d E_{0}/ d R$, displays  a discontinuity at $R=5\,a_0$, but it
is smooth elsewhere (see Fig.~\ref{energiesFig}B). This discontinuity is found
in the first derivatives with respect to $R$ of all the  quantities we
consider. $d E_{0}/ d R$  has a maximum at $R=4.7\,a_0$. From
Fig.~\ref{energiesFig}B (inset and main panel) we see that at first the
shrinking of the inner well increases the ground state energy with an
increasing ``speed''. However, in the migration region the change in the
ground-state energy rapidly slows down: in this region the wavefunction is starting to
spread into the larger outer well, hence moving towards a regime where  $E_{0}$
is almost constant with $R$.

The second derivative of $E_{0}$ with respect to $R$ displays a marked minimum
at $R=4.90\,a_0$ and an infinite discontinuity at $R=5\,a_0$, see
Fig.~\ref{energiesFig}C.

The behaviors of the Coulomb energy $\left\langle U\right\rangle$, and of the
kinetic energy $\left\langle T\right\rangle$ are plotted in the upper panel of
Fig.~\ref{three-entropyFig}, where $\left\langle\dots\right\rangle$ indicates
the ground-state expectation value. For the DIW potential, both display a
maximum located at  $R=4.47\,a_0$   (corresponding to an inner to  outer well
ratio of $0.058$).  The ratio between the  Coulomb and the  kinetic
interactions, Fig.~\ref{three-entropyFig}B,  provides an unambiguous signature
of the migration point $R_{c}$,  whereas  no particular structure emerges from
the visual inspection of both  Coulomb and kinetic energy separately,
Fig.~\ref{three-entropyFig}A.

\subsection{Entanglement}

We calculate the spatial entanglement\cite{Coe:2008:PRB} using the von Neumann
entropy $S$  and the linear entropy $L$,
\begin{align}
& S=-Tr \rho_{red} \log_{2} \rho_{red},\\
& L=Tr(\rho_{red}-\rho_{red}^{2})=1-Tr\rho^{2}_{red}\label{entropies},
\end{align}
where  $\rho_{red}=Tr_{A} |\Psi \rangle  \langle \Psi|$ is the reduced density
matrix found by tracing out the spatial degrees of freedom of one of the two
particles (subsystem `A'
) and $\Psi$ is the ground-state. We consider also the
position space-information entropy $S_{n}$,
\begin{align}
& S_{n}=-\int n(x)\ln n(x)dx,
\label{entropiesb}
\end{align}
where $n(x)$ is the system particle density.

For a pure bipartite state the von Neumann entropy $S$ is the unique function
that satisfies all the entanglement measurement
conditions,\cite{amico:2008:RMP,martin:2007:QIC}  while the linear entropy  $L$
is computationally convenient and quantifies the entanglement in the sense that
it gives an indication of the number and spread of terms in the Schmidt
decomposition of the state.   The position-space information entropy $S_{n}$
can be considered as an approximation to $S$ when off diagonal terms are
neglected \cite{Coe:2008:PRB} and is written in terms of the particle density,
so it could be more easily and directly accessed by experiments.

\begin{figure}[ht!]
\centering
\includegraphics[width=.4\textwidth]{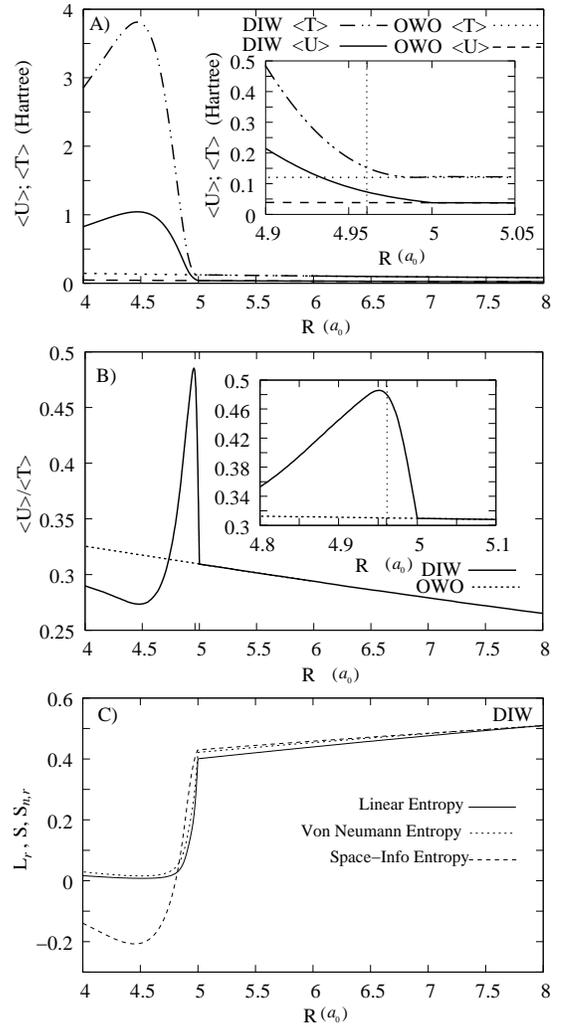}
\caption{Panel A: Coulomb energy, $\left\langle U\right\rangle$,  and kinetic
energy, $\T$, for the DIW  and OWO   potentials as a function of the  driving
parameter $R$. Inset:  as main panel, but in the neighborhood  of the
migration point $R_{c}$, marked by  a vertical dotted line. Panel B: Ratio
between the Coulomb interaction energy and the kinetic energy, $\U/\T$, versus
$R$ for both the DIW and OWO potential. Inset: details of the `migration region'
with the vertical dotted line indicating the point $R=R_c$. Panel C: The von
Neumann ($S$, dotted line), and rescaled linear ($L_r$, solid line)  and
space-information ($S_{n,r}$, dashed line) entropies as  functions of $R$  for
the DIW system. The rescaling was chosen in such a way that $L_{r}$ and
$S_{n,r}$ are equal to $S$ at $R=8\,a_0$. This results in $L_r=3.04L$ and
$S_{n,r}=0.15S_{n}$.}
\label{three-entropyFig}
\end{figure}

  In Fig.~\ref{three-entropyFig}B, the linear, von Neumann,  and position-space
information entropy are plotted as a function of $R$ for the DIW system.  $L$
and $S_n$ have been rescaled so that they have the same value of $S$ at
$R=8\,a_0$. All quantities show the same qualitative behavior, $L$ and $S$
rescaling almost perfectly onto each other. In particular, all quantities show a
{\it non-differentiable point} at $R=5\,a_0$ and present a minimum located in
the same $R$ region. However, the minimum of $S_{n}$ ($R = 4.45\,a_0$) is
nearer to the maximum of $\left\langle U\right\rangle$  than the minima of the
other two entropies ($R= 4.51\,a_0$ for $L$ and $R=4.52\,a_0$ for $S$), and is
more pronounced.

By the Hohenberg-Kohn theorem, \cite{hohenberg:1964:PR} the ground state
particle density uniquely determines all the ground-state properties of the
system, so in principle the ground-state entanglement for this system could be
written as a functional of the density; the overall similarity between $S_{n}$
-- explicitly written as a functional of the density -- and the two
entanglement measures $S$ and $L$ reinforces the idea that pertinent
information can be extracted from the electron density.
As for the DIW system $L$ can be rescaled  very well onto $S$, we will continue
using the computationally convenient linear entropy $L$.

\begin{figure}[ht!]
\centering
\includegraphics[width=.4\textwidth]{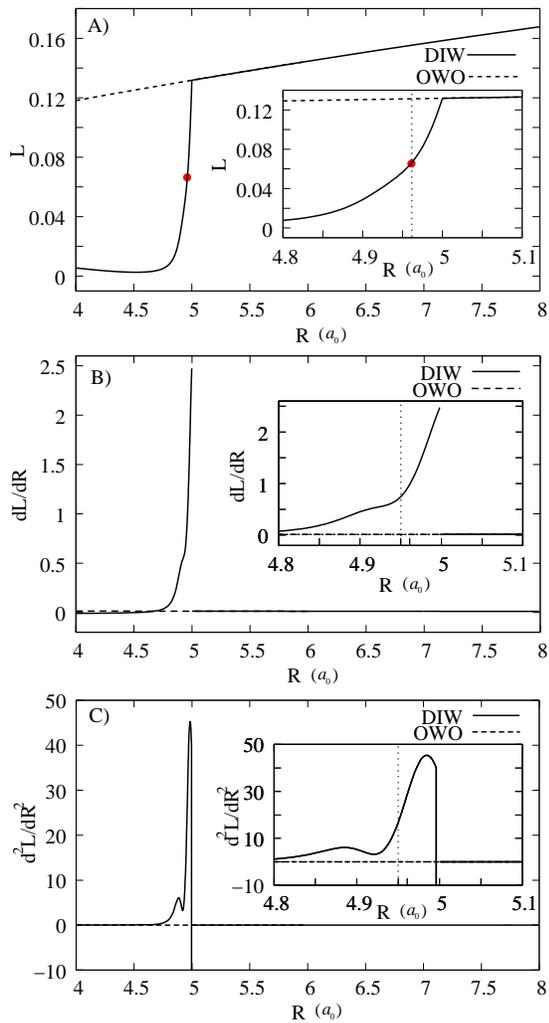}
\caption{Panel A: Linear entropy ($L$)  as a function of R for the DIW and OWO
potentials. The dots indicates the value of $L_{DIW}$ at $R=R_c$. Panels B and C: first and second derivatives  with respect to $R$
of the the linear entropy as a function of $R$.  In all three panels, the inset
represents the respective function around the migration point $R_c$, the latter
being highlighted by the vertical dotted line.}
\label{entanglementFig}
\end{figure}

The linear entropy of the DIW, $L_{DIW}$, and of the OWO system are compared in
Fig.~\ref{entanglementFig}A.
$L_{DIW}$ displays three regions. The first is characterized by a slow
variation in entropy with a shallow minimum at $R= 4.51\,a_0$.  In the second
region ($4.9\,a_0\lesssim R < 5\,a$) the entropy increases very rapidly, and
finally for $R>5a_0$ the entropy increases linearly with $R$. The first
derivative of $L_{DIW}$ (Fig.~\ref{entanglementFig}B) presents  a shoulder-like
structure   connecting the first and
second regions; then, after the boost in the  rate of change of the  entropy,
the derivative has a finite  discontinuity  at $R=5\,a_0$.  The second
derivative of $L_{DIW}$ presents two maxima at $R=4.89\,a_0$ and $R=4.99\,a_0$,
and a minimum at $R=4.92\,a_0$. It has an infinite discontinuity at $R=5\,a_0$

The rate of change of the entropy shown in Fig.~\ref{entanglementFig} is the
result of the competing effects of the confinement strength and of the Coulomb
repulsion. However, since it is the ratio between these two  factors which
governs  the response of the system to a variation of the driving parameter, a
maximum of  $\left\langle U\right\rangle$ corresponds here to a minimum of the
entanglement as these extrema occur when the electrons are most confined
 and hence in an almost factorized state.\cite{Abdullah:2009:PRB}   The
decrease of  $\left\langle U\right\rangle$ is a signature of the wavefunction
spilling into the outer well and, consequently, of an increasing influence of
Coulomb correlations in shaping the wavefunction with a corresponding increase
of the entanglement.  In the migration region (with $\left\langle
U\right\rangle(R_c)$ being approximately $7\%$ of its maximum value), the
wavefunction density is substantially spread within the outer well, and small
variations of $R$ produce large changes in the entanglement.

We note that in \Refon{MarchisioCoeDamicoProc}, a system similar to DIW, with
an  inner well shrinking in width but never disappearing, was studied. In this
a case no discontinuity in any derivative of the relevant quantities were
found.

\subsection{Comparison between the DIW and the OWO potentials}

In Figs.~\ref{energiesFig}A and \ref{entanglementFig}A  $E_{0}$ and $L$ are
plotted for both the OWO and DIW potential.  At $R= R_c$ the many-body
ground-state is bounded to the outer well, and in particular
$E_{0}^{DIW}(R_c)\approx 0.99 E_{0}^{OWO}(R_c)$. In contrast, $L_{DIW}(R_c)$,
which is marked by a dot in Fig.~\ref{entanglementFig}, is approximately {\it
half} of the corresponding entanglement value in the OWO case. We underline
that here the inner well has a {\it finite} depth but the inner to outer well
ratio is only $0.0039$, so, from a geometric point of view, the inner well
should be negligible. However in the region $R_c\le R<5$ the behavior of
$\left\langle U\right\rangle$,  with  $\left\langle U\right\rangle_{DIW} >
\left\langle U\right\rangle_{OWO}$, suggests that the electrons remain strongly
pinned to the inner well region even though  the width of the latter is basically
negligible. As a consequence a very narrow inner well is able to modify the
distribution of the electrons in such a way that their entanglement is highly
and non-linearly reduced. This `pinning property' of the entanglement might
open possibilities of rapidly and efficiently modifying the entanglement in a
nanostructure system.

\section{Ground-state wavefunction and particle-density behavior}

\begin{figure}[ht!]
\centering
  \includegraphics[width=.4\textwidth]{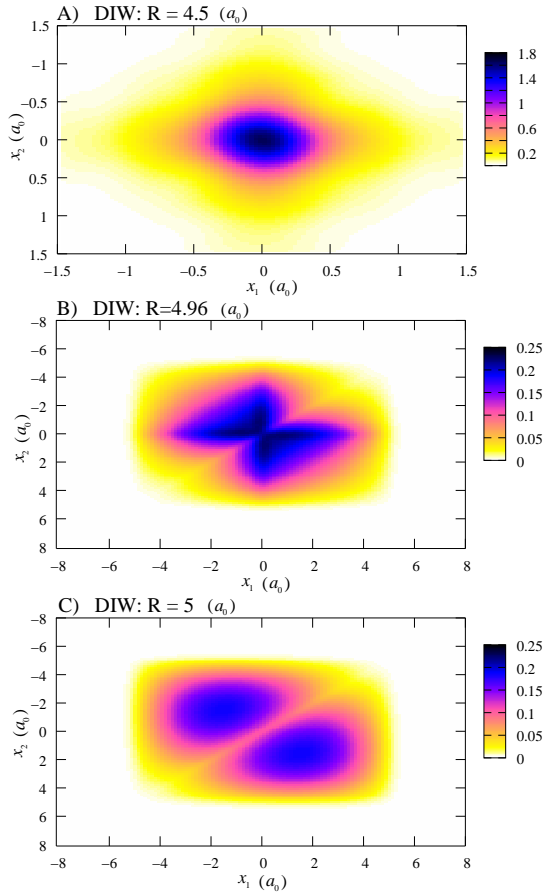}
\caption{(color online) Contour plot of the wavefunction  against the
particles' positions $x_{1}$ and $x_{2}$ for the DIW potential at $R=4.5\,a_0$
($\sim$ maximum of $\left\langle U\right\rangle_{DIW}$ and minimum of $L$, panel A), $R=R_c=4.96\,a_0$ (panel
B), and $R=5\,a_0$ (point at which $W^{iw}_{DIW}=0$, panel C).  }
\label{wf4fig}
\end{figure}

In Fig.~\ref{wf4fig}  the high sensitivity of the system wavefunction to small
changes of the driving parameter in the migration region is explicitly demonstrated.
The figure in fact  shows  the wavefunction contour plots  for $R=4.5\,a_0$
($\sim$ maximum of  $\left\langle U\right\rangle_{DIW}$, panel A), $R=R_c$ (`migration' point, panel B) and $R=5\,a_0$ (value at
which
the inner well disappears, panel C). The wavefunction becomes more and more
confined until $R\approx4.5\,a_0$, for which it displays a single maximum
(panel A).   A further reduction of the inner well width  induces the
wavefunction to leak into the outer well (compare scales on axis of panels A
and B). Around $R\approx R_c$ the wavefunction starts to separate into two
lobes, but remains largest close to the inner well (pinning effect). As $R$
increases beyond $R_c$, the shape of the wavefunction displays two well-defined
lobes, reflecting the effect of the electron-electron repulsion combined with
the diminished confinement strength (panel C). The wavefunction width and
height though remain roughly constant, compare panels B and C. We note that the
wavefunction shape appears to change ``smoothly'' as $R$ increases and, in
particular, no detectable change in the geometry of the wavefunction seems to
take place at $R=5\,a_0$, where the non-differentiable points of the entropy
and energies are both located.

\begin{figure}[ht!]
\centering
  \includegraphics[width=.4\textwidth]{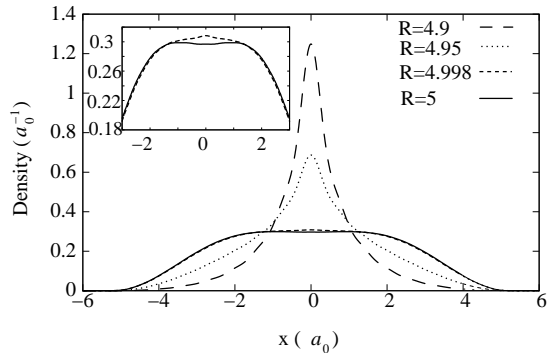}
\caption{Density $n(x;R)$ for the DIW potential plotted against the position
$x$ for four different
values of $R$ (as labeled). Inset: zoom of main panel for
$R\approx5$.}
\label{fig:densityDIW}
\end{figure}

As already mentioned, the particle density $n(x;R)$ should uniquely capture the
system ground-state behavior, so we will now check if the density shape is more
susceptible than the wavefunction to the shrinking and disappearance of the
inner well.
In Fig.~\ref{fig:densityDIW} the density is plotted for various values of $R$.
We see that, as $R$ increases, the height of the central (and only) peak
diminishes. For $R\approx R_c$ the density develops two shoulders  and at
$R=5\,a_0$ the central peak disappears and is
replaced by two peaks which are symmetric around the origin.  This can more
clearly be seen in the inset.
At least for the system at hand,  the pinning from the inner well has a more
clear-cut effect on the shape of the density than on the shape of the
wavefunction, as in particular it determines the presence or absence of a
central peak for the particle density. At difference with the wavefunction, the
change in the number of peaks of the particle density associated to the
disappearance of the central maximum  can then be associated with the
discontinuity in the derivatives of energy and entanglement caused by the
disappearance of the inner well.

\section{Fidelity of the ground-state wavefunction}

The fidelity between two states quantifies their similarity and as such has
been extensively used in quantum information theory.\cite{nielsen}  More
recently the fidelity has been introduced  as a  method for the
characterization of
QPTs:\cite{zanardi:2006:PRE,zanardi:2007:PRL,camposvenuti:2008:PRB} a
signature of QPT is an abrupt change in the wavefunction,\cite{sachdev}  and
this suggests the evaluation of the fidelity  between states across the
critical point as a good choice for the identification of a QPT. Here we will use this method 
to better understand the system behavior in the migration region.

For our system the ground-state fidelity is given by
\begin{equation}
F(R_1,R_2)=|\left\langle
\psi(x_1,x_2;R_1)\left|\psi(x_1,x_2;R_2)\right.\right\rangle|;
\end{equation}
following Eq.~(\ref{eigenfunction}) we  then calculate it as
$F(R_1,R_2)=\left|\sum_{j1,j2=1}^{M}a_{j_1,j_2,n}(R_1)a_{j_1,j_2,n}(R_2)\right|$.
$F(R_1,R_2)$ can be interpreted in two complementary
ways,\cite{Wootters:1981:PRD} and according to this we will consider two
different sets of values for $R_1$ and $R_2$.

In quantum information theory,  the fidelity can be seen as a generalization of
a measure of similarity between two  classical probability
distributions.\cite{nielsen} Let us take $R_1=R_0$, where $\psi(R_0)$  is the
reference state, then the fidelity is the overlap between this initial state
and the wavefunction $\psi(R)$ calculated as $R_2=R$ varies in the parameter
space. The fidelity $F(R_0,R)$ clearly depends on the choice of the reference
state.  The fact that the minimum of the entanglement corresponds to a
quasi-product state (see Fig.~\ref{entanglementFig}, panel A), which evolves 
towards a highly entangled state as $R$ increases, suggests as a natural choice $R_0=4.52\,a_0$, corresponding to the minimum of the linear entropy.

Alternatively, the fidelity can be seen as a geometrical object connected to
the Fubini-Study distance between quantum states,\cite{zanardi:2007:PRL} where
the square distance between infinitesimally close states can be approximated
as $ds_{FS}^{2}\approx 2(1-F)$. In this case  the fidelity is calculated
between two wavefunctions depending on  infinitesimally different parameters,
$\psi(R)$ and $\psi(R+\delta R)$.  At the critical point, where there is an
abrupt change in $\psi$, this function has a minimum and possibly a
discontinuity.
\begin{figure}[ht!]
\centering
  \includegraphics[width=.4\textwidth]{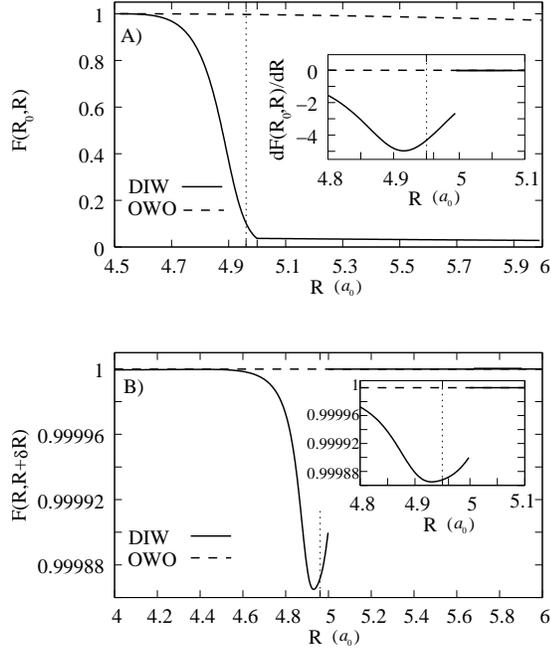}
\caption{Panel A: fidelity $F(R_0,R)$ vs $R$ for $R_0=4.52\,a_0$.  Inset: first
derivative of $F(R_0,R)$ vs $R$. Panel B: fidelity $F(R,R+\delta R)$ vs $R$
with $\delta R = 0.002$. Inset: as for the main panel, but zooming on the
region of the fidelity minimum. The plots refer to the DIW and OWO potential
(as labeled);  the vertical dotted line corresponds to  $R_c$.}
\label{fidelityfig}
\end{figure}
In Fig.~\ref{fidelityfig}, the fidelities $F(R_0,R)$ and $F(R,R+\delta R)$  are
plotted as a function of $R$  (panel A and B, respectively).  $F(R_0,R)$
displays three distinct regimes, in accordance with the behavior of all the
other quantities studied so far. In particular, we see that  for
$4.8\,a_0\lesssim R<5\,a_0$  we have a dramatic decrease of the fidelity: the
wavefunction is rapidly changing from a quasi-product state towards a
triplet-like entangled state\cite{Abdullah:2009:PRB} (compare Fig.~\ref{wf4fig}, panels A and C). The derivative $d
F(R_0,R)/d R$ presents  a  minimum at $R\approx4.92\,a_0$,  near the migration
point $R_c$. For $R>5\,a_0$  the fidelity is almost constant and drastically
reduced, with $F(R_0,R=5)\approx 0.19$: in this region the wavefunction is
nearly orthogonal to the reference state. We note that $F(R_0,R)$ is not
differentiable at  $R=5\,a_0$.

The behavior of $F(R,R+\delta R)$ (Fig.~\ref{fidelityfig}B) shows that the most
significant changes in the wavefunction are confined to the migration region,
around a marked minimum at $R= 4.93$, again very close to $R_c$. $F(R,R+\delta
R)$, which for real wavefunctions and $N$ particles  can be approximated as
\begin{equation}
F(R,R+\delta R)\approx 1- \frac{\delta R^{2}}{2}\!\!\int\!\!\left(\frac{\partial
\psi(x_1\ldots x_N;R)}{\partial R}\right)^2\!\!\!\! dx_1\ldots dx_{N},
\label{F_approx}
\end{equation}
shows a discontinuity at $R=5\,a_0$, in accordance with the discontinuity found
in  the derivatives of all the quantities discussed so far.

\section{Fidelity of the particle density}

For a $1$-particle system with control parameter $R$ we have
that the particle density is $n(x;R)=|\psi(x;R)|^{2}$ so, in this case, the
fidelity may be written in terms of the density as $F(R_{1},R_{2})=\int
\sqrt{n(x;R_{1})n(x;R_{2})} dx$.

We may generalize this to a `density fidelity' by using the density arising
from
$N$-particle systems,
\begin{equation}
n(x,R)=N\int|\psi(x,x_2\ldots,x_n;R)|^2dx_{2}\ldots dx_{n}.
\label{fidinteracting}
\end{equation}
and defining the `density fidelity' as
\begin{equation}
F_{n}(R_{1},R_{2})=\frac{1}{N}\int \sqrt{n(x;R_{1})n(x;R_{2})} dx.
\end{equation}
$F_{n}(R_{1},R_{2})$ has the properties expected from a fidelity, that is $0\le
F_{n}(R_{1},R_{2})\le 1$ and it
measures the overlap between particle densities as the driving parameter $R$ is
varied. We will also demonstrate that
$F_{n}(R_{1},R_{2})$ vanishes if and only if the corresponding wavefunction
fidelity $F(R_{1},R_{2})$ vanishes.

\begin{figure}[ht!]
\centering
  \includegraphics[width=.4\textwidth]{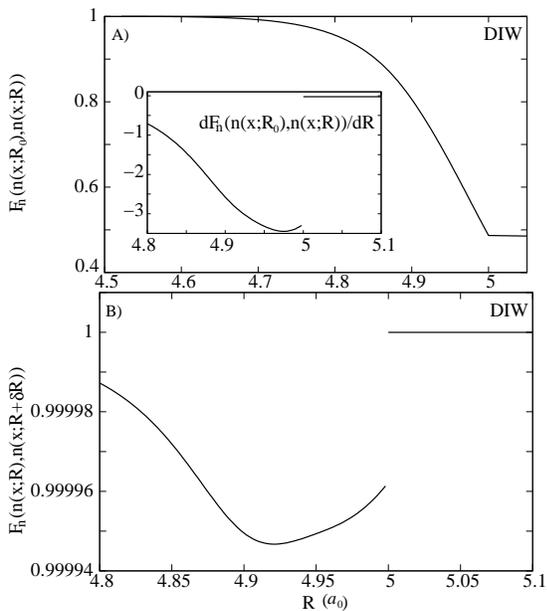}
\caption{ Panel A: Density fidelity $F_{n}\left( R_0,R\right)$,
$R_0=4.52\,a_0$,   for the DIW system, plotted against the parameter $R$.
Inset: Derivative of $F_{n}\left( R_0,R\right)$ with respect to $R$ plotted
against $R$ ($R_0=4.52\,a_0$). Panel B: $F_{n}\left(R,R+\delta R\right)$ with
$\delta R=0.002$, plotted against the parameter $R$ for the DIW system. }
\label{fig:densityfidelity}
\end{figure}

We note that a density fidelity has been proposed for lattice systems and
linked with QPTs in \Refon{Shi-Jian:2009:ChinPhysLett}.
We initially calculate
the density fidelity with respect to $R_0$.   $F_{n}(R_0,R)$  shows a
non-differentiable point at $R=5\,a_0$ corresponding to the disappearance of
the inner well (Fig.~\ref{fig:densityfidelity}A); its derivative in respect to
$R$ is plotted in the inset. We note the similarity between the behavior of
$F_{n}(R_0,R)$ and $F(R_0,R)$ and between their derivatives, the main
difference being that the residual fidelity for $R>\,5a_0$ is larger for the
density than for the wavefunction.

 We show $F_{n}(R,R+\delta R)$, where $\delta R=0.002$, in
Fig.~\ref{fig:densityfidelity} (panel B).  As for the corresponding
wavefunction fidelity, here the discontinuity at $R=5\,a_0$ appears directly in
the `density fidelity'. Again the behavior of $F_{n}(R,R+\delta R)$ and
$F(R,R+\delta R)$  are very similar (compare  Fig.~\ref{fig:densityfidelity}B
with  the inset of Fig.~\ref{fidelityfig}B), with the  density preserving a slightly higher
fidelity at its minimum, which occurs at $R= 4.92\,a_0$

In the DIW system, viewing the density
seems to more clearly and readily display the fast changes in the ground state
properties corresponding to the discontinuity in the derivatives of $E_0$ and
$L$ than viewing the wavefunction (see comments to Figs.~\ref{wf4fig} and
\ref{fig:densityDIW}). This may be due to the lesser formal complexity of the
density, which is always a function of a single position vector -- $x$ in the
present case -- as opposed to the complex many-body wavefunction, a function
of $N$ position vectors whose parameter space is clearly
more difficult to analyse and visualize. In addition the particle density
fidelity is able to predict all the other
notable  features of the wavefunction
fidelity, such as the minimum occurring around $R\approx R_c$. As noted, minima
in the fidelity $F(R,R+\delta R)$ are associated to abrupt changes in the
wavefunction and may signal
the occurrence of a QPT, so, in accordance with
\Refon{Shi-Jian:2009:ChinPhysLett}, our results suggests that the density
fidelity may be used as an alternative to the wavefunction fidelity to
understand brisk changes in the ground state and hence to
study QPTs.
This is in line with the Hohenberg-Kohn theorem which in its simplest form
shows that
for  non-degenerate ground-states, the density uniquely determines the
many-body wavefunction and so all the properties of the
system.\cite{hohenberg:1964:PR}
We point out that the particle density is a much easier quantity to calculate
(and to experimentally access) than the full many-body wavefunction. As such
the use of the fidelity density might become of great help in understanding
phenomena such as QPTs. Similarly its characteristics as highlighted above 
suggest that it could be a useful tool for local sensitivity analysis. 

\subsection{One-to-one correspondence between  vanishing of ground state
particle density fidelity and wavefunction fidelity}
We will now demonstrate the important property that,  for systems with finite
external potentials and ground-states with nodeless spatial wavefunctions, the density fidelity is zero if and only if the ground-state  spatial wavefunction fidelity is zero.

The  nodeless spatial ground-state wavefunction of a time-independent Hamiltonian may always be
taken to be real and positive, so for any two such $N$-particle ground-state
wavefunctions $\psi_1$ and $\psi_2$  we may define the real positive
function
\begin{equation}
f_{1,2}(x)\equiv\int\psi_1\psi_2dx_{2}\dots dx_{N}.
\end{equation}
For any fixed $x$ this defines an inner product, as positive definiteness is
satisfied  by $f_{1,1}(x)=n(x)/N>0$  for {\it finite} external potentials.  The
Cauchy-Schwarz
inequality can then be written as
\begin{eqnarray}
&\int\psi_1\psi_2dx_{2}\ldots dx_{N}\nonumber\\
&\leq \left(\int\left|\psi_1\right|^{2}dx_{2}\dots
dx_{N}\int\left|\psi_2\right|^{2}dx_{2}\dots dx_{N}\right)^{\frac{1}{2}}.
\end{eqnarray}
 integrating both sides with respect to $x$ leads to
\begin{equation}
F \leq F_n
\end{equation}
 and so if  $F_n$ tends to zero, so must $F$.

For a general wavefunction, a fidelity of zero does not imply a density
fidelity of zero as for example two excited state wavefunctions may both be non-zero in
some finite region of space but still be orthogonal.  In addition, if we
compare wavefunctions arising from different forms of inter-particle
interactions, say
attractive and repulsive, then, again, the density cannot always discriminate
between orthogonal wavefunctions.   This can be explicitly seen by considering
the limiting case of
infinite inter-particle attraction or repulsion. Let us consider two particles
in one dimension:
in the case of infinite attraction their wavefunction will satisfy
$\psi_A(x_1,x_2)=0$ if $x_1\ne x_2$, while
for infinite repulsion we have $\psi_R(x_1,x_2)=0$ if $x_1= x_2$,  otherwise
both $\psi>0$. Clearly
we obtain $\int\psi_A\psi_Rdx_{1}dx_{2}=0$.
Let us now consider the single particle densities. In general it will be
$n_R(x)=\int|\psi_R(x,x_1)|^2dx_{1}> 0$.
 To ensure normalization, $\psi_A(x_1,x_2)^2=\phi_1(x_1)^2\delta(x_1-x_2)$,
with $\phi_1(x_1)$ itself normalized, giving a corresponding single particle
density
$n_A(x)=|\phi_1(x)|^2 > 0$. It follows that the related density fidelity $\int
\sqrt{n_R(x)n_A(x)}dx$ is different from zero.

However,  if we assume   the requirements
needed for standard DFT, i.e. ground state and same inter particle interaction,
then we can
argue that the density fidelity can detect orthogonal, nodeless  ground-state
wavefunctions.   The lack of nodes in the ground states means that we can choose a phase so that both our
wavefunctions are never negative. Here a fidelity of zero  corresponds to the
hypothetical situation when the
wavefunctions do not overlap at all.  When the inter particle interaction is fixed this lack
of overlap arises because the wavefunctions are spatially distinct, and so the
densities will not overlap.  Hence for ground-states with nodeless spatial wavefunctions, the density fidelity
is zero if and only if the  spatial wavefunction fidelity is zero.

\section{QPT-like transition (symmetric systems)\label{QPT-like}}
 As pointed out previously, a minimum in $ F(R,R+\delta R)$ may  highlight a
QPT and certainly witnesses a rapid change in the wavefunction. In our case the, 
minimum in $ F(R,R+\delta R)$ observed in Fig.~\ref{fidelityfig} corresponds to
the {\it transition between two separate sets of ground states}; the first set
bounded by the inner and the second set bounded by the outer well. Fig.~\ref{fidelityfig}A shows that this transition is between states that are  {\it almost} orthogonal.  As the width of the inner well is reduced, the energy gap between these set of states reduces: this transition has some of the characteristics of a second-order QPT.

This is apparent when looking at the ground state energy derivatives: $d^2E_0/dR^2$ presents in this region a marked minimum, which in turn corresponds  to an inflection point in the energy first derivative. If this were a full-fledged
QPT, this inflection point would have a vertical tangent, and hence the minimum
in $d^2E_0/dR^2$ would become a divergency.

\begin{figure}[ht!]
\centering
  \includegraphics[width=.4\textwidth]{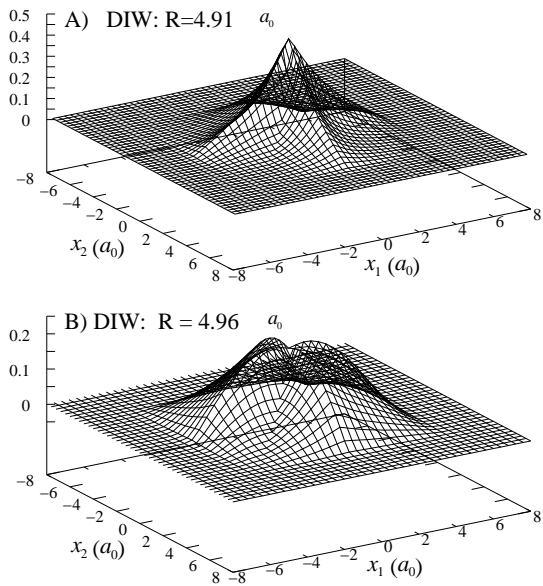}
\caption{ Ground-state wavefunction plotted against the particles' positions
$x_{1}$ and $x_{2}$ for the DIW potential at $R=4.91\,a_0$ , panel A, and
$R=R_c=4.96\,a_0$, panel B. }
\label{wf491fig}
\end{figure}

As discussed in  Refs~\onlinecite{Wu:2006:PRA} and \onlinecite{Wu:2004:PRL}, a second-order QPT should be
signaled by a corresponding structure in the first derivative of the
entanglement.
The first  derivative of the entanglement entropy  presents indeed a structure
(a shoulder) whose width can be defined by the first maximum-minimum structure
in $d^2L/dR^2$, i.e $4.89\,a_0\le R\le 4.92\,a_0$ (see
Fig.\ref{entanglementFig}, panels B and C): this shoulder indeed frames the
region of the minimum of $d^2E_0/dR^2$. The bulk of the wavefunction change should occur
in the region of the minimum of $ F(R,R+\delta R)$:
 in Fig.~\ref{wf491fig} we then present the wavefunction at $R=4.91\,a_0$
(panel A) and  $R=R_c=4.96\,a_0$  (panel B).  The plots confirm a quite
substantial change in the wavefunction,  which smears over the upper well as
$R$ increases, changing from a single, pointed  peak towards a two-lobe
geometry.

As for the case of a finite-size system which would undergo a QPT
in the thermodynamic limit,\cite{Osterloh:2002:nature} the transition we
observe in the wavefunction occurs over a (small) parameter region and  slightly away from the expected `critical'
value of the driving parameter, i.e.,  for $R\lesssim R_c$.

\section{Origin of the discontinuities observed at $R=w$\label{sec}}

It was demonstrated in \Refon{Wu:2006:PRA,Wu:2004:PRL} that a discontinuity in
the first (second) derivative of the ground-state energy with respect to the
driving parameter -- a signal of a QPT --  may correspond to a discontinuity in
the (derivative of the) ground-state entanglement.

What we observe in the present case at $R=w$ is instead a discontinuity in the
{\it same order} derivatives of the ground-state energy and entanglement.
Moreover, all  the other quantities under study, such as $\left\langle
U\right\rangle$ or $S_n$,   are non-differentiable  at the same point.
A similar situation  was speculated for the limit $p\to\infty$ in
\Refon{Abdullah:2009:PRB}. Here we would like to clarify the origin of the
discontinuities we observe.

First of all we can extract from the fidelities important information on the
ground state wavefunction behavior at $R=w$: the continuity of $F(R_0,R)$ shows
that the ground-state wavefunction is {\it  continuous}, on average, at $R=w$
(see Fig.~\ref{fidelityfig}A). However the discontinuity of $\partial
F(R_0,R)/\partial R|_{R=w}$ indicates a  discontinuous derivative for the wave
function at the same point (inset of  Fig.~\ref{fidelityfig}A). 
The discontinuity of $\partial  \psi(\bm{x};R)/\partial R $ at  $R=w$  is confirmed by the discontinuity of
$F(R,R+\delta R)$  at the same point, see Eq.~(\ref{F_approx}).

To understand the above picture we consider the  Hamiltonian   for a general
potential

\begin{equation}
H(R)=H_{0}+\sum_{j}V(x_j,R),\label{ham0}
\end{equation}
where  $H_{0}={ T} +{ {U}}$ is independent from  the driving parameter $R$, $
T$ is the kinetic energy, and $ {U}$ is the electron-electron interaction. From
the Hellmann-Feynman theorem we have
\begin{equation}
\frac{d E(R)}{d R} = \sum_{j}\left\langle\psi(\bm{x};R)\left|\frac{\partial
V(x_{j};R)}{\partial R}, \right|\psi(\bm{x};R)\right\rangle,
\label{HF}
\end{equation}
where $\bm{x}=(x_1,\dots x_N)$ represents the coordinates of the $N$ particles.
Consequently, if $\partial V(x_{j};R)/\partial R$ is discontinuous with respect
to the parameter $R$, then this discontinuity could propagate to  $d{E}/d{R}$.

We note that in the case of a  first order QPT, the discontinuity in
$d{E}/d{R}$ should arise from the wavefunction, and not from the potential. In
the present case, while the fidelity indicates a continuous wavefunctions, not
only $d{E}/d{R}$, but also all the other quantities used as indicators for a
QPT, present a point of non-analyticity at $R=w=5\,a_0$. It is hence necessary
to understand how a discontinuity in   the potential would affect the other
quantities of interest, and in particular if, in contrast with the situation in
Refs.~\onlinecite{Wu:2004:PRL}, \onlinecite{Wu:2006:PRA},    it would produce
discontinuities in the \textit{first} derivative of \textit{both} energy and
entanglement entropies.

By considering the time-independent Schr\"odinger equation associated to
Eq.~(\ref{ham0}) we can write

\begin{equation}
-\frac{ T\psi(\bm{x};R)}{\psi(\bm{x};R)}-\frac{{U}
\psi(\bm{x};R)}{\psi(\bm{x};R)} + E(R) =\sum^{2}_{j=1}V(x_{j};R).
\label{dft1}
\end{equation}

  Eq.~(\ref{dft1}) is well-defined for a nodeless ground state wavefunction and
can be seen as a family of equations labeled by the continuous parameter $R$.
Eq.~(\ref{dft1}) shows  that if the potential is discontinuous only at a set of
points of measure zero then, at most, it may only directly cause $\psi$ and/or
$ T\psi$ to be discontinuous on that same set of points.  Hence, for
\textit{finite} discontinuities, these discontinuities will not propagate to
any integrated quantities such as expectation values.
We then continue by assuming that $\psi$ and ${T}\psi$ are, at worst,
discontinuous  over a set of points of measure zero.
We differentiate Eq.~(\ref{dft1}) with respect to $R$, and use Eq.~(\ref{HF})
to obtain

\begin{align}
&-\frac{1}{\psi(\bm{x};R)} T\frac{\partial{\psi(\bm{x};R)}}{\partial{R}} +
\frac{\left[
T\psi(\bm{x};R)\right]}{\psi^2(\bm{x};R)}\frac{\partial{\psi(\bm{x};R)}}{\partial{R}}\nonumber\\
& =\sum_{j}\left[ \frac{\partial V(x_{j};R) }{\partial
R}-\left\langle\psi(\bm{x};R)\left|\frac{\partial V(x_{j};R)}{\partial R}
\right|\psi(\bm{x};R)\right\rangle\right].
\label{dft2}
\end{align}
Finite discontinuities in the potential may mean that its derivative with
respect to $R$ will comprise delta functions, and hence that, unless accidental
cancellations occur, these discontinuities will propagate to
$\left\langle\partial{V(x_{j};R)}/\partial{R}\right\rangle$.  Let us assume
that they are such that
$\left\langle\partial{V(x_{j};R)}/\partial{R}\right\rangle$ is discontinuous
at  $R=\tilde{R}$. Then, on the right-hand side of Eq.~(\ref{dft2}) we have two
discontinuous functions with respect to $R$, but as the second term does not
depend on $\bm{x}$, the right-hand side is actually discontinuous  at
$(\bm{x};\tilde{R})$ {\it for all or almost all values of $\bm{x}$} since no
accidental cancellations  can hold for all $\bm{x}$. This means that the left
hand side of Eq.~(\ref{dft2}) will present the same  discontinuities. As
$\psi(\bm{x};R)$ and $ T\psi$ are at least continuous almost everywhere with
respect to $\bm{x}$ at $(\bm{x};\tilde{R})$,   this implies that
$\partial{\psi(\bm{x};R)}/{\partial{R}}$  has indeed to be discontinuous at
$(\bm{x}; \tilde{R})$  for all  or almost all $\bm{x}$ and hence the first
derivative in respect to $R$ of any functional  of $\psi$ will be in general
discontinuous at $R=\tilde{R}$. This is exactly what we observe.

In Appendix A1 we  illustrate these points by explicitly analyzing the effect of
the finite discontinuity  at the point $(x=0;R=w)$ in  $V_{DIW}$.

In the next section we will instead   consider a counter-example for which,
due to an accidental cancellation,
$\left\langle\partial{V(x_{j};R)}/\partial{R}\right\rangle$
-- and hence all first derivatives in respect to $R$ -- remains continuous even
in the presence of discontinuities in $V(x_{j};R)$ similar to the ones of  the
DIW potential.

\section{Core-shell to double well potential}

\begin{figure}[ht!]
\centering
  \includegraphics[width=.4\textwidth]{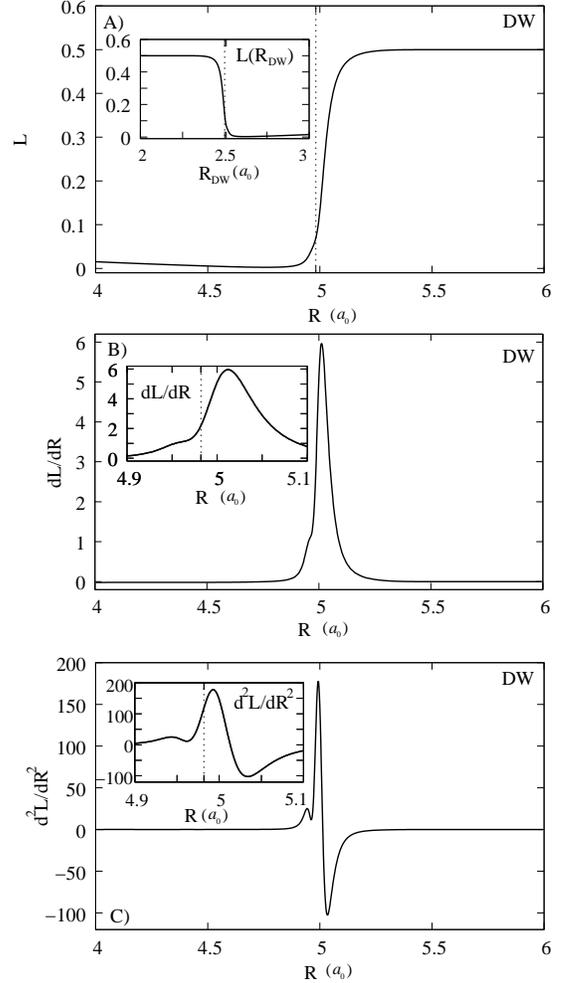}
\caption{Upper panel: Entanglement entropy $L$ plotted against $R$ for the
potential of Eq.~(\ref{eq:doublewell}) and $w=5a_0$. Inset: same as main panel
but plotted in respect to $R_{DW}$ for an easier comparison with
\Refon{Abdullah:2009:PRB}. Middle panel and lower panel: $dL/dR$  and
$d^2L/dR^2$, respectively,  vs $R$ for the potential of
Eq.~(\ref{eq:doublewell}) and for  $w=5\,a_0$. The insets present the
transition region and the vertical dotted line the migration point
$R_{c}=4.98\,a_0$  for the  potential Eq.~(\ref{eq:doublewell}).}
\label{fig:doublewell}
\end{figure}

In \Refon{Abdullah:2009:PRB} it was speculated that, in the rectangular-like
potential limit,
 the observed sharp transitions in energies and entanglement
would display non-analyticities  as the potential changes from a core-shell
structure to a
double well potential.
The behavior of the entanglement and its derivatives in this limit is shown in
Fig.~\ref{fig:doublewell}.

\begin{figure}[ht!]
\centering
\includegraphics[width=.4\textwidth]{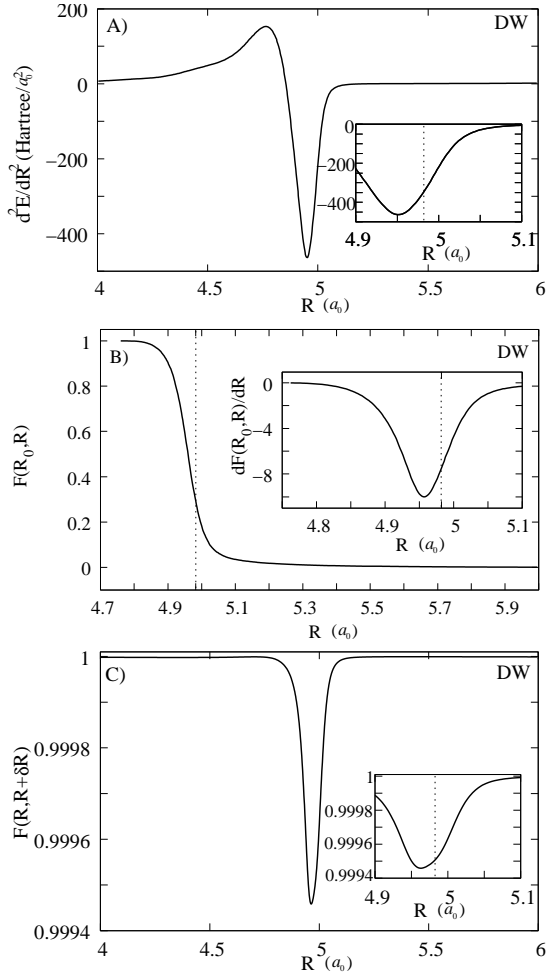}
\caption{Panel A, B and C: $d^2E_0/dR^2$, $F(R_0,R)$  and $F(R,R +\delta R)$
for the DW  potential with respect to $R$. The insets in panel A and C present
the details of the transition region; the inset in B depicts the first
derivative of $F(R_0,R)$ with respect to $R$ and with $R_0=4.76a_0$.
} \label{SPfidFig} \end{figure}

For the DW potential Eq.~(\ref{eq:doublewell}), the plot of   $F(R_0,R)$
(Fig.~\ref{SPfidFig}B, with $R_0=4.76a_0$ the minimum of $L$ for {\it this}
system)  confirms that the reference state, almost factorized and bounded to
the inner well, is practically orthogonal to the triplet-like ground state
reached {\it after the transition to the double well
potential}.\cite{Abdullah:2009:PRB} However a QPT-like transition occurs only
when the  ground state bounded to the inner well migrates to the outer well, see
the shoulder in  $dL/dR$ and the minimum of $d^2E_0/dR^2$ in
Figs.~\ref{fig:doublewell}B and \ref{SPfidFig}A, respectively. The subsequent
transition to a double-well potential  merely further isolates the two lobes of
the wavefunction from each other.
The sharp increase in entanglement which corresponds to this further change,
does not then signal any further QPT-like point, as confirmed by the absence of
additional structures in   $d^2E_0/dR^2$ and $F(R,R +\delta R)$ (see
Fig.~\ref{SPfidFig}A and Fig.~\ref{SPfidFig}C).

Interestingly, in contrast with the DIW case, the discontinuity at $R=w$ of
$\left\langle\partial{V(x_{j};R)}/\partial{R}\right\rangle$ is   {\it
accidentally removed} in this system.   As a result the derivatives with
respect to $R$ of the various quantities of interest, as well as $F(R,R+\delta
R)$, {\it do not present discontinuities} (see Figs.~\ref{fig:doublewell} and
\ref{SPfidFig}).  Details are given in Appendix A2.

\section{Asymmetric potential}

\begin{figure*}[ht!]
\centering
\includegraphics[width=1\textwidth]{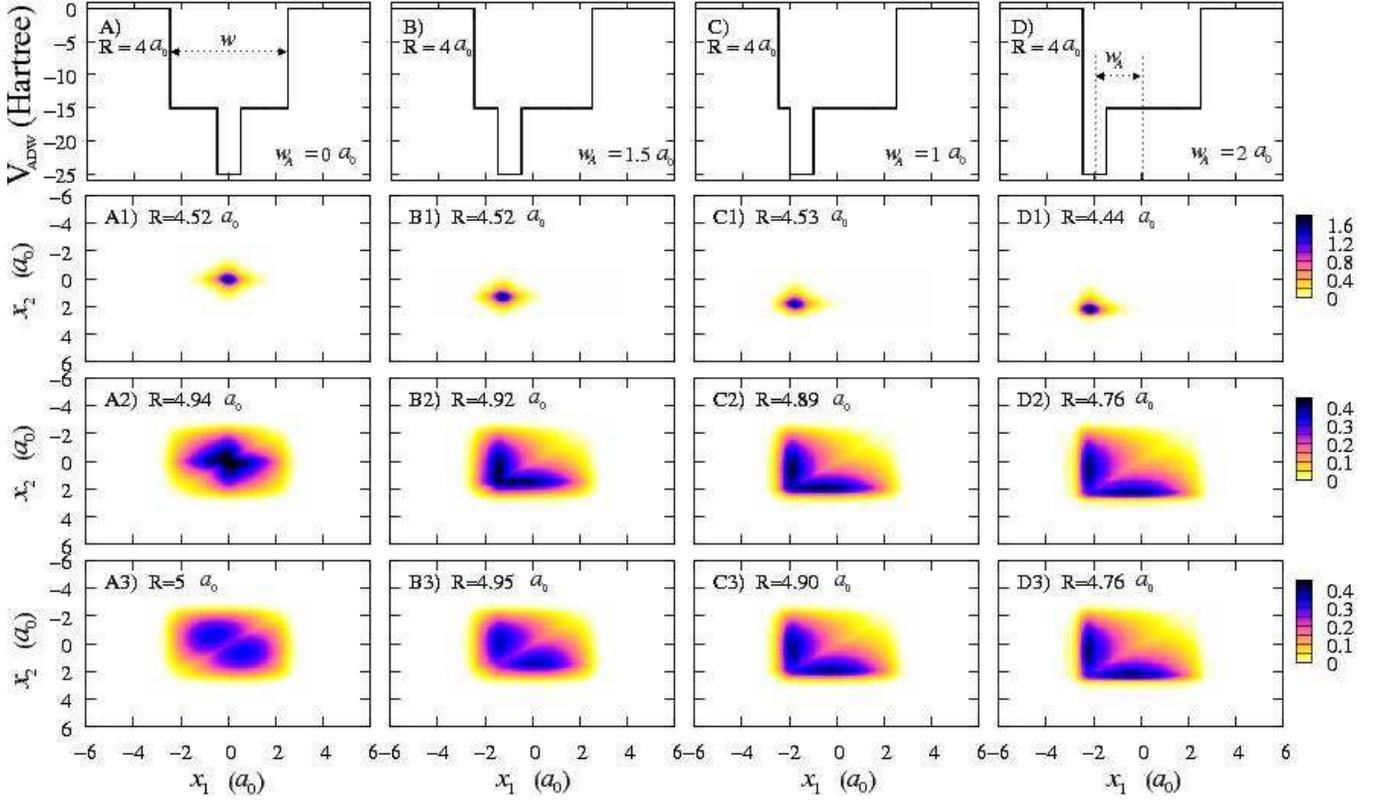}
\caption{(Color online) Panels B to D: Asymmetric Disappearing inner Well (ADW) potential  $w_a=1\,a_0$ Vs the coordinate $x\equiv x_1$. Panel B and C, same as above but for   $w_a=1.5\,a_0$  and $w_a=2\,a_0$. Panel A: symmetric case ($w_a=0\,a_0$).
Panels A1 to D1: Contour plots of the wavefunction against the
particles' positions ($x_1$, $x_2$) at the minimum of $L(R)$. Panels A2 to D2: same as above but in correspondence of the `migration' point. Panels A3 to D3:  same as above but in correspondence of the maximum of the entropy. } 
\label{AsymPot}
\end{figure*}

We will now consider the possibility for the position of the inner well, with respect to the symmetry  axis of the outer well,
   to be shifted toward left (see Fig.~\ref{AsymPot}, upper panels) and we will refer to these asymmetric potentials as Asymmetric Disappearing inner Well (ADW) potentials.  
In Fig~\ref{AsymPot}D, the inner well position is left-shifted by a quantity $w_a=2\,a_0$, where $w_a$ is the distance between the symmetry axes of the inner and outer well.  In panels~\ref{AsymPot}C  and \ref{AsymPot}B,  $w_a=1.5\,a_0$  and $w_a=1\,a_0$, respectively, while the system described in panel~\ref{AsymPot}A, which represents the symmetric case $w_a=0\,a_0$,  is taken as benchmark.   The width of the inner well follows Eq.~(\ref{Wiw}) with $w=5\,a_0$,  shrinking from $W^{iw}=1\,a_0$ for $R=4\,a_0$  to  $W^{iw}=0\,a_0$ for $R=5\,a_0$. The width of the outer well is assumed constant, $W^{ow}=w$. The depth of the wells is $V^{ow}_0=-15$ Hartree and s $V^{iw}_0=-10$ Hartree for the outer and inner well discontinuity (barrier) rispective. Henceforth, the letters A, B, C, D are used to indicate  the four systems whose potential is depicted in the respective panels of  Fig~\ref{AsymPot}.

\begin{figure}[ht!]
\centering
\includegraphics[width=.4\textwidth]{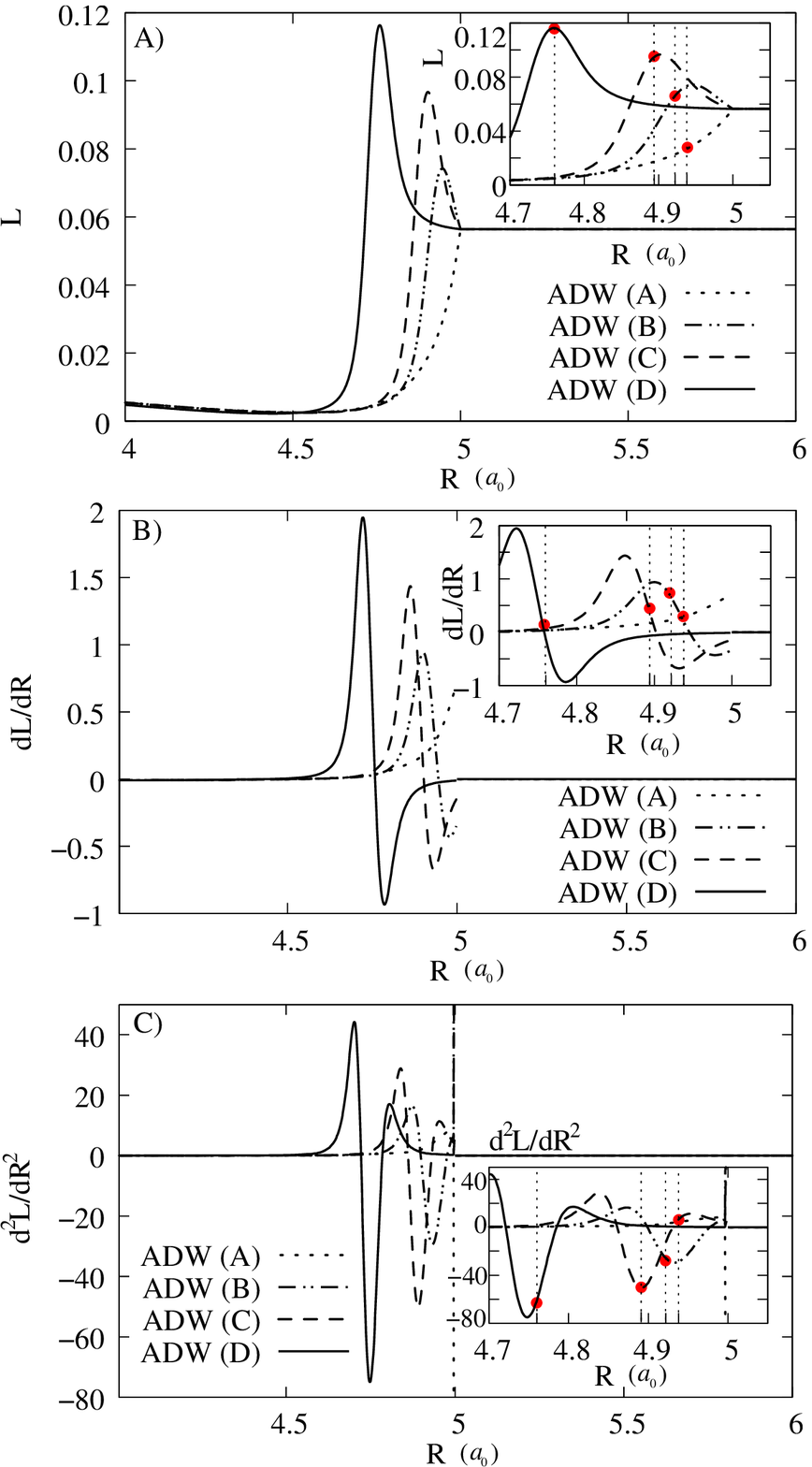}
\caption{Panel A, B and C: $L(R)$, $dL/dR$ and $d^2L/dR^2$ plotted as a function of the driving parameter $R$   for the four potentials depicted in Fig.~\ref{AsymPot}. The  region of major variations of these quantities  are detailed  in the  inset inside the respective panels. The dotted vertical lines highlight the migration point for each potential.} 
\label{entcomp}
\end{figure}

The entanglement (linear entropy) and its derivatives, $dL/dR$ and $d^2L/dR^2$, are plotted in the three panels of   Fig.~\ref{entcomp}, respectively.  For $R\approx 4\, a_0$, the many-body wavefunction is well-confined inside the inner well. Thus the entanglement of systems A to D shows there approximately the same behaviour, independently of the position of the inner well with respect to the symmetry axis of the outer well. The second row of panels in Fig~\ref{AsymPot}, namely A1,B1,C1 and D1, shows  the contour plots of the two-body wavefunction with respect to the particles' coordinates $x_1$ and $x_2$ in correspondence of the  minimum of the linear entropy.
 With the exception of the case D ( Fig.~\ref{AsymPot}D1 ), the contour plots are very similar  where the wavefunction is significantly different from zero. This, in turn, is reflected in a similar position of the minimum of the linear entropy. In fact, using the subscripts  A, B, C and D to indicate the four potentials, we have that the  minima of $L(R)$ are at $R_{A}=4.52\,a_{0}$, $R_{B}=4.52\,a_{0}$, $R_{C}=4.53\,a_{0}$,  and  $R_{D}=4.44\,a_{0}$. 
 
\begin{figure}[ht!]
\centering
\includegraphics[width=.4\textwidth]{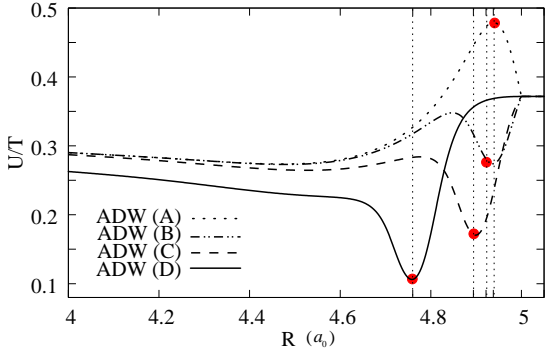}
\caption{Ratio $\U/\T$  (Coulomb energy over kinetic energy) for the four potential represented in Fig.~\ref{AsymPot} with respect to $R$.  The dotted vertical lines highlight the migration point for each case A, B, C and D. } 
\label{Fidcomp2}
\end{figure}

\begin{figure}[ht!]
\centering
\includegraphics[width=.4\textwidth]{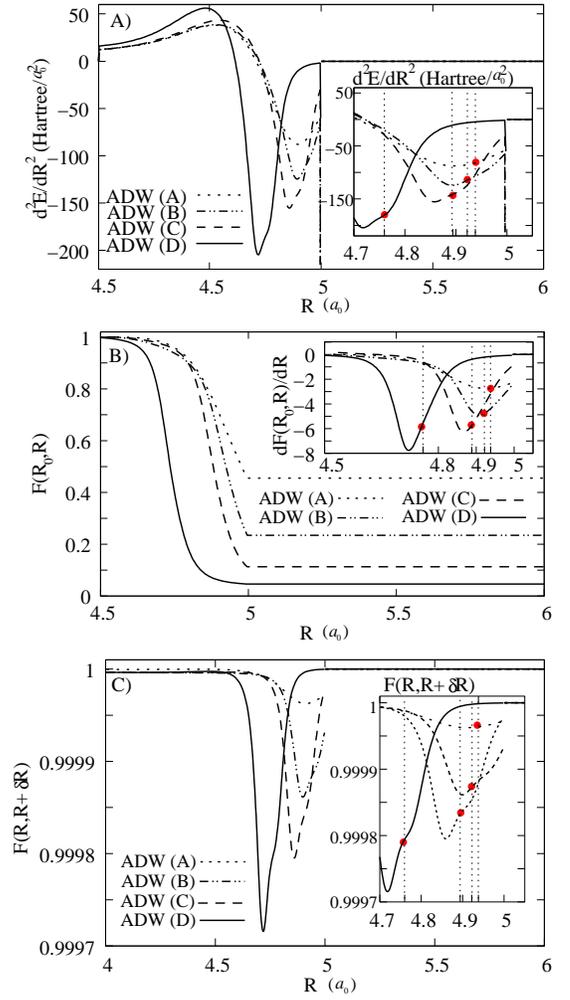}
\caption{Panel A, B and C: $d^2E/dR^2$, $F(R_0,R)$ and $F(R,R+\delta R)$ plotted as a function of the driving parameter $R$   for the four potentials depicted in Fig.~\ref{AsymPot}. $R_{0}$ is the value of the parameter at which $L(R)$ reaches its minimum for each potential in  Fig.~\ref{AsymPot}, where $\delta R=0.002\,a_0$ for all four cases.  The inset represent the region around the migration point for the panels $A$  and $C$, while in panel $B$ the inset depicts the derivative of $F(R_{0}, R)$ with respect to $R$.  The dotted vertical lines highlight the migration point for each potential.} 
\label{Fidcomp}
\end{figure}
 
\begin{figure}[ht!]
\centering
\includegraphics[width=.4\textwidth]{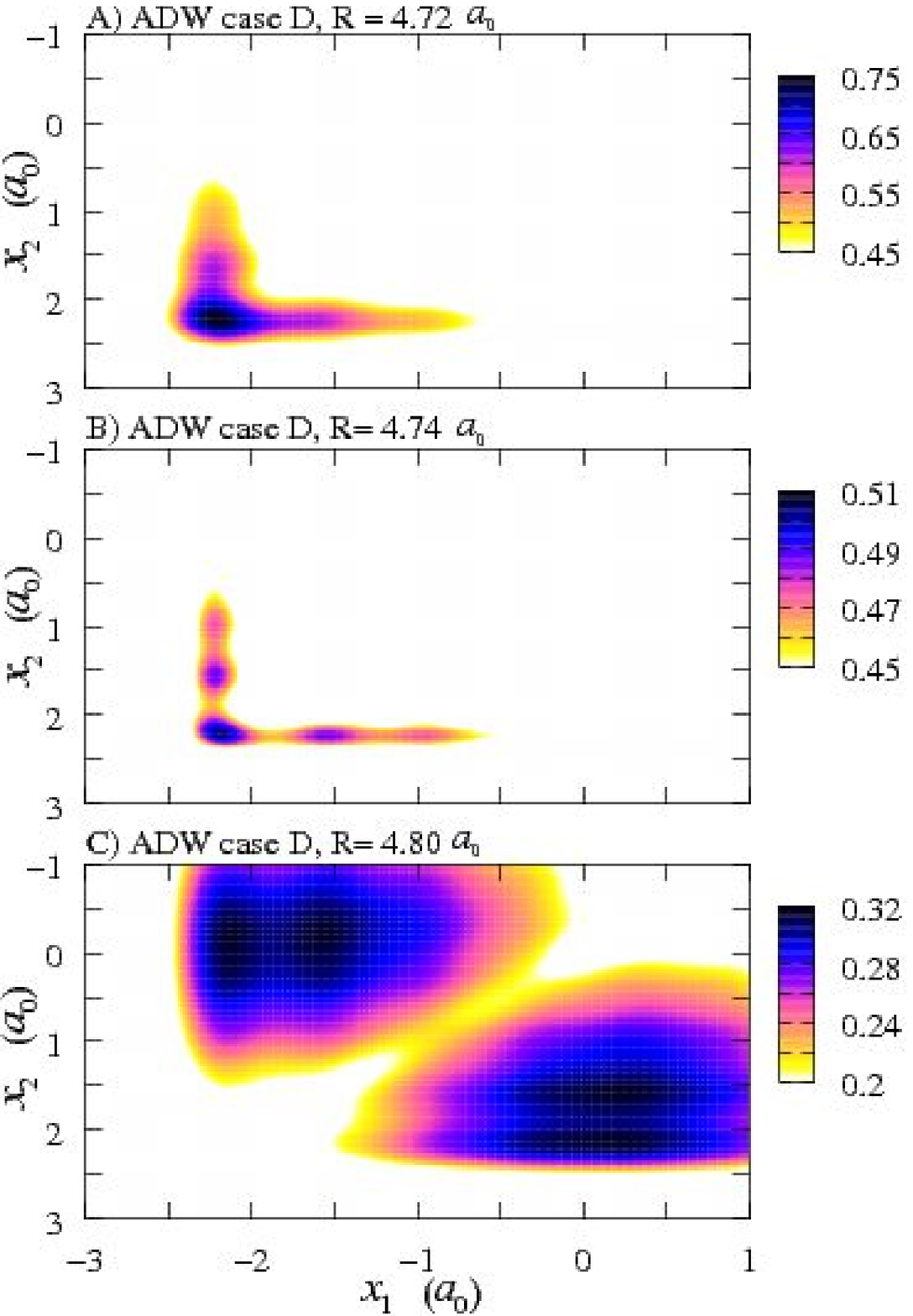}
\caption{Panel A, B, C:  Asymmetric Disappearing inner Well (ADW) case D. Contour plots of the wavefunction against the
particles' positions ($x_1$, $x_2$) at $R=4.72\, a_0$ (minimum of $d^2E/dR^2$ and $F(R,R+\delta R)$) $R=4.74\, a_0$ and $R=4.80\, a_0$ respectively } 
\label{sisD}
\end{figure}

As $R$ increases,  the ground-state approaches  the migration point. Consequently, the particles become progressively bounded by very different confinement potentials: symmetric for A, and increasingly asymmetric from B to D. As the ground state wavefunction `spreads' in the upper  well, the larger is $w_a$, the more the wavefunction is distorted by a combination of the pinning effect due to the finite size of the  inner well and the presence of the left hand side potential barrier of the outer well.  The  third row of panels in Fig~\ref{AsymPot} highlights the increasingly marked asymmetry in correspondence of the migration points, located at $R_{D}=4.76\,a_{0}$,  $R_{C}=4.89\,a_{0}$,  $R_{B}=4.92\,a_{0}$ and $R_{A}=4.94\,a_{0}$ (panels from A2 to D2).

The most interesting feature is, however, the marked maximum of the entanglement entropy for the three asymmetric potentials. In particular, the maximum of $L_{D}(R)$ is {\it twice as large} as the maximum value reached in the symmetric case. For the symmetric case, the maximum entanglement is reached at $R=5\,a_{0}$. The other maxima are located at $R_{D}=4.76\,a_{0}$,  $R_{C}=4.90\,a_{0}$ and $R_{B}=4.95\,a_{0}$, hence  the maximum of the entropy is reached at a value of $R$ which become closer and closer to the migration point as the asymmetry increases. For case D,  the two values coincide. These and the appearance of the entanglement maxima may be understood by looking at  Fig.~\ref{AsymPot}A3 to Fig.~\ref{AsymPot}D3, which display the contour plot of the wavefunction at the entanglement maxima: (i) the pinning effect localizes most of the wavefunction at the location of the inner well, while (ii) the reduced distance of outer well walls from the inner well in the asymmetric potentials further reduces the possibility for the wavefunction to significantly expand in the outer well. Due to  (i) and (ii), as the asymmetry increases,  the wavefunction becomes more strongly separated into two lobes and more strongly localized in two narrow peaks. These two characteristics imply that more information could be learned about the second particle position once the first is measured and hence means a larger entanglement with respect to the symmetric case. In particular,  for case D, both (i) and (ii) are maximized at the migration point, which then corresponds to the maximum of the entanglement.     

Using the same argument as in Sec.~\ref{sec}, it  is easy to show that the expectation value of the first derivative  with respect to $R$ of the potential $ADW$, is discontinuous at $R=5\,a_{0}$. It follows that for this value of $R$, the function  $L(R)$  is non-differentiable for all four cases, see Fig.~\ref{entcomp}B. Finally, we note that, for each asymmetric potential, the minimum of the second derivative of the entropy $d^2L/dR^2$  occurs approximately in correspondence of the migration point; while the  extreme point of the ratio $\U/\T$ signals the migration point for both symmetric and asymmetric  geometries of the confinement, see Fig.~\ref{Fidcomp2}.

\section{QPT-like transition (asymmetric systems)\label{QPT-likeasy}}
Similarly to the symmetric case (see Sec.~\ref{QPT-like}), also for asymmetric potentials  the minimum of  $d^2E/dR^2$ and of the  fidelity  $F(R,R+\delta R)$ is reached at the same value of the driving parameter: $R_{D}=4.72\,a_{0}$,  $R_{C}=4.86\,a_{0}$,  $R_{B}=4.89\,a_{0}$ and $R_{A}=4.91\,a_{0}$, see  Fig.~\ref{Fidcomp}A and Fig.~\ref{Fidcomp}C. In addition,  the minimum of $dF(R_0,R)/dR$  also occurs at very similar values of $R$, namely  $R_{D}=4.72\,a_{0}$,  $R_{C}=4.87\,a_{0}$,  $R_{B}=4.90\,a_{0}$ and $R_{A}=4.91\,a_{0}$.
Notably the minima of $d^2E/dR^2$ and $F(R,R+\delta R)$ become the more pronounced the  more asymmetric the system is.  

The shoulder structure of $dL/dR$ observed in symmetric systems in correspondence to the migration point (and discussed in Sec.~\ref{QPT-like}) develops now into a maximum-minimum structure (middle panel in Fig.~\ref{entcomp}). Here  the maximum corresponds to the minimum of $d^2E/dR^2$ fairly precisely for systems B (maximum of $dL/dR$ at $R=4.90$, minimum of $d^2E/dR^2$ at $R=4.89$) and precisely for systems C and D. This coincidence of $d^2E/dR^2$ and $dL/dR$ extrema has been shown to signal QPTs in some systems\cite{Gu:2003:PRA,Chen:2006:NJP}. We then expect the system at hand to display related characteristics, and in particular a rapid evolution between two (almost) orthogonal states. This is confirmed by Fig.~\ref{Fidcomp}B showing that when the driving parameter is swept across the minimum structure of $d^2E/dR^2$ ($4.5\alt R\alt 5$) the system rapidly evolves between two almost orthogonal states.

The minimum of $d^2E/dR^2$ and of $F(R,R+\delta R)$ becomes the more pronounced the  more asymmetric the system is, indicating a more marked QPT-like behavior. We expect then a {\it qualitative} change of the wavefunction to occur on an even smaller parameter range, i.e. when sweeping the driving parameter just across the minimum itself. This is indeed the case: by looking at the wave-function countour plots for system D, we see that by sweeping the driving parameter in the range $4.71\alt R \alt 4.74$ the wavefunction evolves from having a single maximum positioned at the inner well, to having additional, marked, lateral maxima. These display the increasing importance of Coulomb correlations over confinement (compare Fig.~\ref{sisD} panels A and B).

Interestingly in the asymmetric systems the `migration' point marks a secondary structure (a shoulder) on the right of the minimum of $F(R,R+\delta R)$ {\it and} $d^2E/dR^2$. In this region the wavefunction maximum located above the inner well  disappears, and, due to Coulomb correlations, a valley develops between the wavefunction lobes (Fig.~\ref{sisD}C).

The very narrow parameter region between the minimum of $F(R,R+\delta R)$ {\it and} $d^2E/dR^2$ and the `migration' point represents the crossover from confinement to Coulomb correlations as the leading term in shaping the wave-function characteristics. In this region the entanglement is extremely sensitive to variations of the external parameter, going from  its almost minimum to its maximum value (seeFig.~\ref{entcomp}). This very strong sensitivity could be exploited to envisage entanglement 'switches'.

\section{Conclusions}

We studied a set of systems of two electrons  confined by a potential which evolves
from  a core-shell to a single well potential. 
For symmetric confining potentials (DIW system)
as the  driving
parameter $R$ increases, the inner well width is reduced while the width of the
outer well increases. The system ground state is initially bounded to the inner
well with a wavefunction in a quasi-product state (entanglement minimum and
Coulomb energy maximum). In the region around the point where the two
electrons' ground-state migrates into the outer well, the energy and  the entanglement
entropies are continuous and differentiable. However, here various QPT markers 
display a behavior similar to the one of  a second-order QPT:  in particular the
second derivative of the ground state energy, as well as the fidelity
$F(R,R+\delta R)$, displays a minimum, and the first derivative of the ground
state entanglement presents a structure around this minimum. 
we then
associate this parameter region to a `QPT-like' transition. The wavefunction
indeed   undergoes here a drastic change, quickly
evolving from a quasi-product towards a triplet-like state.  We showed that a
similar QPT-like transition characterizes also the evolution from a core-shell
to a double-well potential described in \Refon{Abdullah:2009:PRB}.

We compared the results for the DIW system  with a benchmark system where only
the outer  well is present, and noticed that a very narrow inner well has a
strong pinning effect on the entanglement (and fidelity) of the system. In
particular for this symmetric system, the entanglement may be reduced by half by the presence of an inner
well, {\it even when the ground state is already bounded to the outer well}.

We also consider  systems (ADW) with the position of the core well being {\it asymmetric} with respect to the shell. We show that, in the 'migration' region, their sensitivity to small variation of the driving parameter is even greater than for the symmetric case, with the entanglement displaying a sharp maximum at the `migration' point and a value as high at twice the maximum value of a system without the inner well and {\it about four times the value of a corresponding DIW system}.
This property might be exploited to create entanglement switches by moving the position of the inner well with respect to the outer well confining potential.

The maximum in the entanglement of ADW systems derives from a combination of the 'pinning effect' by the inner core and of the asymmetric confinement (in respect to the pinning center) provided by the outer shell. This interplay results in the high probability of the second particle to be found in a more sharply defined small region of space, once the position of the first is measured, hence the increase in  entanglement.
In this region the analysis of QPT markers, and in particular the coincidence of the minimum of the energy second order derivative  with the maximum of the entanglement entropy first order derivative, confirms a QPT-like behavior for the system wavefunction. This transition occurs in the region where, as the wavefunction expands from the inner to the outer well, Coulomb correlations starts to substantially modify its shape, and is due to the interplay between confinement by the core well, Coulomb interactions, and confinement by the outer well.

We demonstrated  that a potential which is characterized by a driving parameter
$R$ and has a finite discontinuity, even just at a single point $(\tilde{x},
\tilde{R})$, may induce a discontinuity in the {\it derivative} of the
ground-state wavefunction at $\tilde{R}$ for any (or almost any) value of
$x$. This in turn  induces a non differentiability at $\tilde{R}$, with
respect to $R$, of any functional of the ground state and in particular -- and
in contrast with QPT signatures -- of the {\it same order} derivatives of energy
and entanglement. This is the case for the DIW potential described in this
work: here the non-analytic  behavior of ground state energy and entanglement,
which is reminiscent of  the one encountered in QPT, derives  instead in a
non-trivial way from the {\it finite discontinuity} at a {\it single point} of
the confining potential. To underline the peculiarity of this connection we
also presented a counter-example in which similar discontinuities in the
potential {\it do not} transfer to other quantities.
This is the case of the  rectangular-like  limit of the potential considered in
\Refon{Abdullah:2009:PRB}, where exact cancellations occur, and hence the
ground state wavefunction and all related quantities, such as the entanglement
and energy, remain differentiable at any $R$.

We presented  a detailed analysis of the {\it particle density fidelity}, and
showed that, for the ground state, this quantity provides similar  information
to the wavefunction fidelity, but may be calculated directly from the more
accessible particle density. In particular we demonstrate that,  for  ground-states with nodeless spatial wavefunctions, {\it the particle density fidelity is zero if and only if the wave
function fidelity is zero}.

Our results suggest that the entanglement  of two particles confined in a
controllable core-shell structure would display QPT-like characteristics as the
ground states `migrates' from the inner well to the outer shell, including a well
defined minimum of both the wavefunction and particle density fidelities. In the corresponding 
narrow driving parameter range our calculations
demonstrate a high sensitivity of the entanglement to variations of the core well size 
and/or position.

Finally our results show that the particle density fidelity may complement
or perhaps even replace the traditional wavefunction fidelity as a diagnostic tool of
 the ground state of  those systems for which the particle density may be
experimentally accessed.

\appendix

\section{Effect of the finite discontinuity of the confining potential}
\subsection{Core-shell to single well potential (DIW potential)}
We note from  Eq.~(\ref{pot_na1}) and Eq.~(\ref{pot_na2}) that
$V_{DIW}(x;R)$ undergoes linear transformations (contraction of the inner
well, dilatation of the upper well)   which  are continuous  everywhere in $R$
with the exception of  the point $(x=0;R=w)$. Treating the potentials
$V_{DIW}(x;R)$ as a family of square integrable functions (this condition
is verified for  $R<\infty$), the usual distance can be defined and in
particular  we have that $D\big(V_{DIW}(x;w-\delta
R),V_{DIW}(x;w)\big)=\sqrt{\int\left|V_{DIW}(x,w-\delta R) -
V_{DIW}(x,w)\right|^{2}dx}$ goes to zero continuously as $\delta R \rightarrow
0$. This means that the disappearance of the inner well at $R=w$ marks the
transition between two potentials which differ over a set of measure zero.

We now consider the expectation value of the derivative of our potential,
$\langle {\partial V_{DIW}(x;R)}/{\partial R}\rangle$, which is directly
related to the derivative of the ground-state energy, see  Eq.~(\ref{HF}).

Of the two components of $V_{DIW}$, see Eq.~(\ref{analyticalpot}), only the
derivative of $V^{iw}$ will contribute
to a possible discontinuity at $(x=0;R=w)$, as
the outer well is simply widening as $R$ increases. Its derivative is given by

\begin{align}
&\frac{\partial V^{iw}(x;R)}{\partial R} =\nonumber\\ &-
\frac{V_{0}}{2}\left\{ \delta\left( x +
\frac{w-R}{2}\right)\theta\left( -x +
\frac{w-R}{2}\right)\right.\nonumber\\ &+ \left.\theta\left( x +
\frac{w-R}{2}\right)\delta\left( -x + \frac{w-R}{2}\right)\right\}.
\label{deripot} \end{align}
We can calculate  $\langle {\partial V^{iw}(x;R)}/{\partial R}\rangle$ with the
help of the property
\begin{equation}
 \int_{-\infty}^{+\infty} n(x;R)\theta(\pm x+x_0)\delta(\mp x+x_0)dx=
c_{x_0}n(\pm x_0;R)
\end{equation}
where  $c_{x_0}=1$ for $x_0>0$, $c_{x_0}=0$ for $x_0\le 0$.
By using the symmetry $n(x;R)=n(-x;R)$ and considering $R<w$,
we find

\begin{align} &\left.\langle\frac{\partial
V^{iw}(x;R)}{\partial R}\rangle\right|_{R<w}=
-\frac{V_{0}}{2} n(W^{iw}/2;R). \label{pot1} 
\end{align}
On the other hand, for $R>w$, we obtain
\begin{align} &\left.\langle\frac{\partial
V^{iw}(x_{j};R)}{\partial R}\rangle\right|_{R>w}=0.
 \label{pot1deriv} 
\end{align}

For $R\to w^{\pm}$, the above equations give

\begin{equation}
\left.\langle\frac{\partial
V^{iw}(x;R)}{\partial R}\rangle\right|_{R\to
w^+}-\left.\langle\frac{\partial
V^{iw}(x;R)}{\partial R}\rangle\right|_{R\to w^-} =\frac{V_{0}}{2} n(0;w).
\label{potdisc}
\end{equation}

Here $n(0;w)$ is  different from zero, and hence
the  expectation value of the derivative of the potential with respect to $R$
is
discontinuous at $R=w$.

As a consequence, the derivative of the  wavefunction with respect to $R$ is
discontinuous  at $(\bm{x}; R=w)$ for all or almost all $\bm{x}$, and this
explains the discontinuities observed in $F(R,R+\delta R)$ and $F_n(R,R+\delta
R)$ and in the derivatives in respect to $R$ of entanglement, energies,
$F(R_0,R)$, and $F_n(R_0,R)$.
\subsection{Core-shell to double well potential (DW potential)}
Similarly to the DIW potential, the potential in  Eq.~(\ref{eq:doublewell})
displays a finite discontinuity in respect to $R$ at $(x=0; R=w)$.

For $R<w$, Eq.~(\ref{eq:doublewell}) is similar to the DIW potential except
that in  Eq.~(\ref{eq:doublewell}) the  outer well decreases as the inner well
does. For $R \geq w$, Eq.~(\ref{eq:doublewell}) describes two separated wells
which move further apart and decrease in width as $R$ increases.

By repeating calculations similar to the ones done for $V_{DIW}$ we obtain

\begin{equation}
 \left.\langle\frac{\partial
V_{DW}(x_{j};R)}{\partial R}\rangle\right|_{R\to w^\pm}
=-\frac{V_{0}}{2}\big[n\big(0;w)+n\big(w;w)],
\end{equation}
which shows that   {\it no discontinuity} appears in the limit $R\to w^{\pm}$
in the hypothesis that the wavefunction, and hence $n(x,R)$ is a continuous
function of $R$: in this case, even if $\partial V_{DW}/\partial R$ does contain
delta function-type discontinuities, there is no discontinuity in
$\langle\partial V_{DW}/\partial R\rangle$ and hence  the derivative of the
wavefunction is continuous as well as the derivatives of the other functions
discussed in this paper.
This picture is confirmed by  Fig.~\ref{fig:doublewell} where calculations
performed directly with the square-well potential in Eq.~(\ref{eq:doublewell})
show a steep gradient of the entanglement at $R=w=5\,a_0$, but not a
discontinuity in its derivative.

We remark that this cancellation of the discontinuity is accidental and is due
to the fact that the depth of the inner well is the same as the height of the
barrier between the double-well structure.  The discontinuity  reappears if
this symmetry is lifted.

\bibliography{biblioEDITpaper}

\end{document}